# Brain activity vs. seismicity: Scaling and memory

Sumiyoshi Abe [1,2,3,4] and Norikazu Suzuki [5]

[1] *Department of Physics, College of Information Science and Engineering, Huaqiao University, Xiamen 361021, China*
[2] *Department of Physical Engineering, Mie University, Mie 514-8507, Japan*
[3] *Institute of Physics, Kazan Federal University, Kazan 420008, Russia*
[4] *ESIEA, 9 Rue Vesale, Paris 75005, France*
[5] *College of Science and Technology, Nihon University, Chiba 274-8501, Japan*

**Abstract.** The brain activity and seismicity share a remarkable similarity. The Gutenberg-Richter law describing a power-law relation between the frequency of earthquake occurrence and released energy has its counterpart in the brain activity of a patient with epilepsy, that is, the distribution of fluctuations of the voltage difference measured by electroencephalogram (EEG) also obeys a Gutenberg-Richter-like power law. The similarity in the distributions, however, does not directly tell if the processes underlying these intermittent phenomena are also similar to each other. Here, a new simple method is presented for quantitative evaluation of (non-)Markovianity and is applied to the processes of released energy in seismicity and fluctuation of the voltage difference in EEG data. It is shown that the process in seismicity is almost memoryless, whereas that in EEG has long-term memory. ☐

The Gutenberg-Richter law [1] is placed at the position of central importance in seismology. It manifests how the distribution of earthquake energy is exotic. Let $N(E)$ be the frequency of occurrence of earthquakes with released energy $E$ during a certain period of time, e.g., annual frequency. Then, the Gutenberg-Richter law states that $\log N(E) \sim A - bM$, where $b\,(>0)$ and $A$ are constants, and $M$ is magnitude defined in terms of $E$ as $\log E = 11.8 + 1.5M$ (for significant earthquakes with $M > 6.5 \sim 7.0$, another definition called moment magnitude denoted by $M_W$ is often used). This implies that the distribution of earthquake energy itself asymptotically decays as a power law with no characteristic scales, in marked contrast to ubiquitous laws of the exponential type.

During the last two decades, similarities of the specific brain activity to seismicity have repeatedly been studied [2-7] (see also Ref. [8] for a general reading). Of particular interest is in electroencephalogram (EEG) signals of patients with epilepsy. An EEG records the values of the voltage difference, $V$, between two selected electrodes located on the surface of a patient's scalp. (The intracranial EEG will not be treated in the present work.) The time series of the absolute value of $V$, between two points on the surface of the scalp is intermittent like earthquake energy, and the distribution of $|V|$ decays as a power law: $P_{EEG}(|V|) \sim |V|^{-\alpha}$ with $\alpha$ being a positive exponent, like the Gutenberg-Richter law.

Now, for the claimed similarity, yet it is necessary to go beyond the discussion about the forms of the distributions. In this respect, we emphasize that the microscopic dynamics generating these two processes are still largely unknown, although some attempts at grasping their nature have been made in the literature. For example, see Refs. [9-12] for the brain activity and Refs. [13-15] for seismicity. Here, "microscopic dynamics" implies a map from the value of the state variable of one event/signal to that of the next, provided that the state variable may be released energy in seismicity and signal amplitude in EEG. Under such a circumstance, an important step toward extracting information on the dynamics is to reveal the property of memories, i.e., temporal correlations.

In this article, we study the properties of memories in the brain activity and seismicity. A traditional method of investigating memory is based on the autocorrelation function of relevant variables [16]. However, since characterizing (non-)Markovianity is a major issue itself, any novel approach may be welcome. Therefore, we wish to contribute also to this point, here. Thus, we present a new method for quantitatively evaluating (non-)Markovianity of a given empirical time series, which is actually applicable to diverse systems and phenomena, not limited to the brain activity and seismicity. Using this method, we show how these two phenomena are different from each other: process of the released energy in seismicity is almost memoryless (i.e., Markovian), whereas that of the EEG signals contain long-term memory (i.e., non-Markovian).

To examine independencies of our result from geographical regions and patients, we analyze two different data sets each for seismicity and EEG signals. For seismicity, they are the ones taken from southern California provided by the Southern California Earthquake Data Center (available at http://scedc.caltech.edu) and from Japan provided by National Research Institute for Earth Science and Disaster Resilience (available at http://www.hinet.bosai.go.jp). The data set referred to as "**Seismicity-1**" is from southern California; the total time interval analyzed is between 00:25:08.58 on January 1, 1984 and 23:11:07.13 on December 31, 2018, containing the earthquake event type and local geographical type; the region covered is 30.09483°N–37.9855°N latitude, 113.15717°W–121.76000°W longitude, and -2.6–51.1 km in depth, and "**Seismicity-2**" is from Japan; the total time interval analyzed is between 00:02:29.62 on June 3, 2002 and 15:54:32.84 on December 11, 2006; the region covered is 17.956°N–48.060°N latitude, 120.119°E–155.520°E longitude, and 0–681 km in depth. On the other hand, for the EEG signals, they are the ones recorded from two different patients with epilepsy at Boston Children's Hospital that are available through Refs. [17,18] in CHB-MIT Scalp EEG Database (available at https://physionet.org/cgi-bin/atm/ATM). The data set referred to as "**Brain-1**" is from an 11-year-old patient; a part of the data set named "chb01_18" containing the EEG records of the voltage difference between two points on the surface of the scalp labeled $F_{P1}$ and $F_2$ (of the total 21 points where electrodes are set), and "**Brain-2**" is from a 3-year-old patient; a part of the dataset named "chb13_62" containing the EEG records of the voltage difference between two points on the surface of the scalp labeled $T_7$ and $FT_9$.

All of these four are adjusted to contain the events and signals of the common total number 591465. Seismicity-1 and Seismicity-2 respectively include, as the disastrous main shocks, the Baja California Earthquake with $M_W 7.2$ occurred at 22:40:42.36 on April 4, 2010 and the Fukuoka-ken Seiho-oki Earthquake with $M_{JMA}7.0$ occurred at 10:53:40.32 on March 20, 2005 ("*JMA*" indicates that this value is determined in the Japan Meteorological Agency scale). In Brain-1, the first signal ($n=1$) is adjusted to the 176536th signal in the data set in order for a single strong epileptic seizure to be included, whereas it is the 200001st signal for three seizures to be included in Brain-2.

Fig. 1A shows a part of the time series of the released energy in Seismicity-1, and the corresponding normalized distribution of the energy obtained from the whole Sseismicity-1 is presented in Fig. 1B. A scaling region, which is a straight line in the log-log plot, is observed in Fig. 1B, demonstrating the validity of the Gutenberg-Richter law. On the other hand, the time series of the absolute value of the voltage difference in Brain-1 and the corresponding normalized distribution are plotted in Figs. 2A and 2B, respectively. The EEG signals consist of 256 samples per second with 16-bit resolution by the use of the internationally standardized 10-20 system. The resulting time series is seen to be intermittent like in seismicity, and the distribution in Fig. 2B has a scaling region, exhibiting its similarity to Fig. 1B. Therefore, making the absolute value of the voltage difference correspond to the earthquake energy in seismicity, we in fact observe validity of the Gutenberg-Richter-like power-law distribution in Fig. 2B.

The following point should be noted. The physical significances of the scale in the conventional time, e.g., 1 sec, in seismicity and in the brain activity are largely different from each other. In the case of Seismicity-1, it takes 35 years for 591465 earthquakes to occur, whereas 591465 EEG signals are sampled for about 2310.41 sec. This is why *n*, which is referred to as *event time* in seismicity and *signal time* in EEG, plays a distinct role for description of the dynamics.

Particular attention should be focused on the chronologically ordered subsequence, $\{x_1, x_2, ..., x_\mathcal{N}\}$, of an empirical process that yields a pure power-law distribution, where $x_k$ ( $k = 1, 2, ..., \mathcal{N}$ ) is the $k$th value of the event/signal size. It can be extracted from the whole sequence by lower and upper threshold settings for event/signal size (see Figs. 1B and 2B). The subsequences thus obtained in the brain activity and seismicity are precisely responsible for a claimed similarity between the two scaling phenomena. For $m = 1, 2, 3, ...$, the value $x_k$ ( $k = 1, 2, ..., \mathcal{N} - m$ ) is regarded as the $k$th realization of a random variable, $X$. Consider another random variable $X'$, the $k$th realization of which is the shifted one: $x_{k+m}$ ( $k + m = m + 1, m + 2, ..., \mathcal{N}$ ). Suppose $X$ and $X'$ jointly yield a normalized distribution, $F(x, x')$. (Here and hereafter, the case of continuous variables is considered for convenience.) Then, the variable, $r = Q(X, X')$ with $Q$ being a certain bivariate function, is distributed according to $P(r) = \langle \delta(r - Q(X, X')) \rangle$, where the angle brackets denote the expectation value with respect to the joint distribution, $F(x, x')$. If the process is strictly Markovian, then $X$ and $X'$ are independent each other, and the joint distribution is factorized: $F(x, x') = f(x) g(x')$, where $f(x)$ and $g(x')$ are the distributions of $X$ and $X'$, respectively. *Until this stage, these two distributions are arbitrary.* Now, of interest for us is the case when both $f(x)$ and $g(x')$ are pure power-law distributions in finite intervals:

$$f(x) = \frac{C}{x^\alpha} \quad (\varepsilon \leq x \leq \Lambda), \qquad g(x') = \frac{C'}{(x')^{\alpha'}} \quad (\varepsilon' \leq x' \leq \Lambda'). \tag{1}$$

In these expressions, $\alpha$ and $\alpha'$ are positive exponents. $\varepsilon$ and $\varepsilon'$ are positive lower thresholds, whereas $\Lambda$ and $\Lambda'$ are upper thresholds, as mentioned above. $C$ and $C'$ are the normalization constants given by $C = (1-\alpha)/\left(\Lambda^{1-\alpha} - \varepsilon^{1-\alpha}\right)$ ($\alpha \neq 1$), $1/\ln(\Lambda/\varepsilon)$ ($\alpha = 1$), and similarly for $C'$. Furthermore, what is relevant to the subsequent analysis of ours turns out to be the specific case when $g(x') = f(x')$. That is,

$$\alpha = \alpha' \equiv \alpha^*, \qquad \varepsilon = \varepsilon', \qquad \Lambda = \Lambda', \tag{2}$$

where the common exponent, $\alpha^*$, is chosen in such a way that $g(x') = f(x')$ is established for empirical data to a good approximation. A bivariate function we consider here is a simple one: $Q(X, X') = X'/X$, which obviously respects the scale invariance, $Q(\lambda X, \lambda X') = Q(X, X')$, with $\lambda$ being a positive constant. Then, the distribution, $P(r)$, of

$$r = \frac{X'}{X} \tag{3}$$

is found to be given as follows:

(i) $\alpha^* \neq 1$;

$$P(r) = \begin{cases} \dfrac{1-\alpha^*}{2\left[(\varepsilon/\Lambda)^{1-\alpha^*}-1\right]^2}\left[\dfrac{1}{r^{\alpha^*}}-\dfrac{(\varepsilon/\Lambda)^{2-2\alpha^*}}{r^{2-\alpha^*}}\right] & (\varepsilon/\Lambda < r \le 1) \\[2ex] \dfrac{1-\alpha^*}{2\left[(\Lambda/\varepsilon)^{1-\alpha^*}-1\right]^2}\left[\dfrac{(\Lambda/\varepsilon)^{2-2\alpha^*}}{r^{2-\alpha^*}}-\dfrac{1}{r^{\alpha^*}}\right] & (1 < r < \Lambda/\varepsilon) \end{cases}, \qquad (4)$$

(ii) $\alpha^* = 1$;

$$P(r) = \begin{cases} \dfrac{1}{\left[\ln(\varepsilon/\Lambda)\right]^2}\dfrac{\ln\left[r/(\varepsilon/\Lambda)\right]}{r} & (\varepsilon/\Lambda < r \le 1) \\[2ex] \dfrac{1}{\left[\ln(\Lambda/\varepsilon)\right]^2}\dfrac{\ln\left[(\Lambda/\varepsilon)r\right]}{r} & (1 < r < \Lambda/\varepsilon) \end{cases}. \qquad (5)$$

$P(r)$ in Eq. (4) behaves like a double power-law distribution. If a process yielding a pure power-law distribution is Markovian, then the variable in Eq. (3) should obey the above distribution.

Before proceeding, we wish to make a couple of comments. Firstly, if $r$ is shifted as $r-1$, then the variable becomes analogous to *return* in finance, where $X$ is the price of an asset [19]. Secondly, a joint distribution given by the product of two identical pure power-law distributions with $\Lambda \to \infty$ has been discussed in Ref. [20] for a purpose different from ours. There, the authors have examined how the so-called $q$-Gaussian distribution can approximately be obtained for the variable, $|X - X'|$, not our $r$, for real seismicity as well as its self-organized criticality model.

To quantitatively evaluate the difference between $P(r)$ and the distribution calculated from the real data $P_{data}(r_i) \left[ = P_{EQ}(r_i), P_{EEG}(r_i) \right]$, the values of $P(r)$ at the corresponding discrete points $\{r_i\}_{i=1,2,...,N}$ are taken. After being normalized again, the discrete distribution is written anew as $P(r_i)$. Then, taking the scaling nature into account, we propose to quantify such a difference by making use of the "normalized" squared logarithmic distance

$$d^2(P, P_{data}) = \frac{1}{N} \sum_{i=1}^{N} \left( \log P(r_i) - \log P_{data}(r_i) \right)^2, \qquad (6)$$

where we have introduced the "normalizing" prefactor $1/N$ in order to eliminate the data-size dependence. The exponent $\alpha^*$ in Eq. (2) should be used in Eq. (6).

The results on Seismicity-1 and Brain-1 are shown in Figs. 3 and 4, respectively. Together with Seismicity-2 and Brain-2, the values of threshold, the exponents, and the normalized squared logarithmic distance are presented in Tables I, II, and III. In Table III, we see that the sequences of the EEG signals contain long-term memory in terms of signal time, whereas those of earthquake energy in event time do not. In fact, the values of the normalized squared logarithmic distance in Eq. (6) are very small already for the minimal shift, $m = 1$, in seismicity and are comparable to $m = 10000$ in the cases of EEG.

In conclusion, we have quantitatively shown by using a new method presented here that the process of EEG signal size in the brain activity possesses a long-term memory but that of the released energy in seismicity is almost memoryless, although both of them exhibit the scaling phenomena similarly to each other. Thus, the dynamics governing the process in the brain activity may be considered to be operating near the onset of chaos, at which the maximum Lyapunov exponent [21] is vanishingly small and the system remembers its initial condition for a very long duration of time without mixing [22], in marked contrast to the released energy in seismicity that should be strongly chaotic without memory. In other words, state transitions in seismicity (i.e., the changes of the released energy) are caused by a temporally-local mechanism, whereas those in the brain activity contain temporal nonlocality exhibiting the complexity of the underlying dynamics.

The authors have been supported in part by a Grant-in-Aid for Scientific Research from the Japan Society for the Promotion of Science (No. 16K05484). S.A. gratefully acknowledges the supports by a grant from National Natural Science Foundation of China (No. 11775084), the program of Fujian Province, China, and the program of Competitive Growth of Kazan Federal University from the Ministry of Education and Science, Russian Federation. N.S. is indebted to a Grant-in-Aid (A) of College of Science and Technology, Nihon University.

———————————————

# Figures

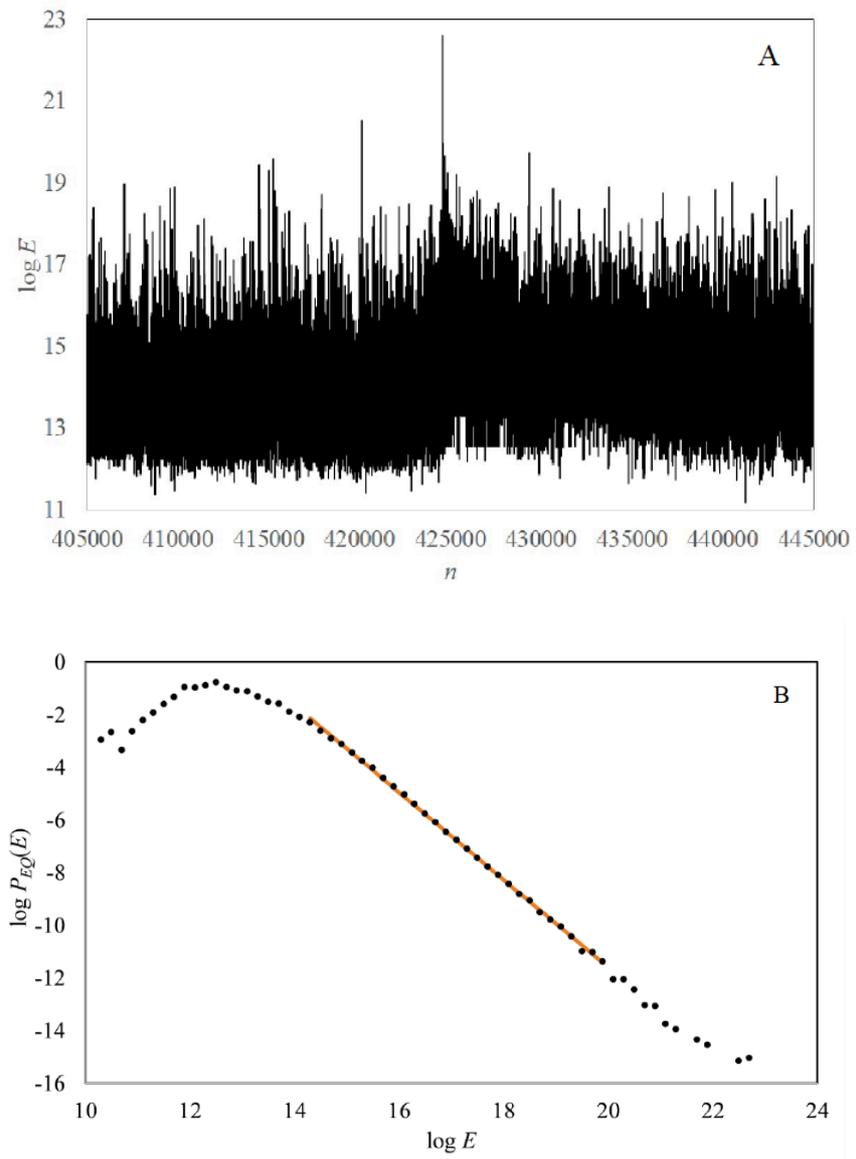

FIG. 1

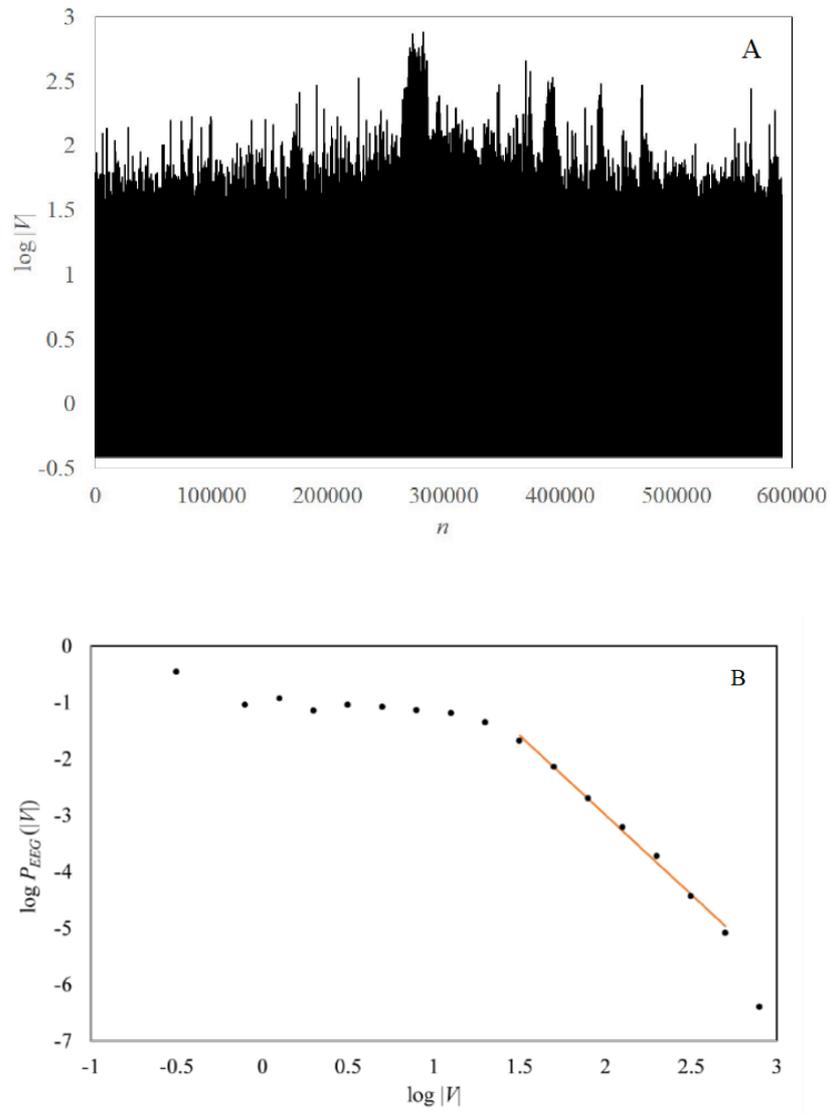

FIG. 2

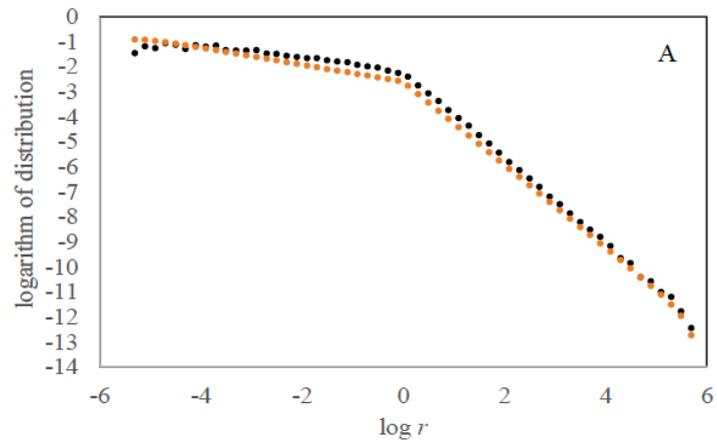

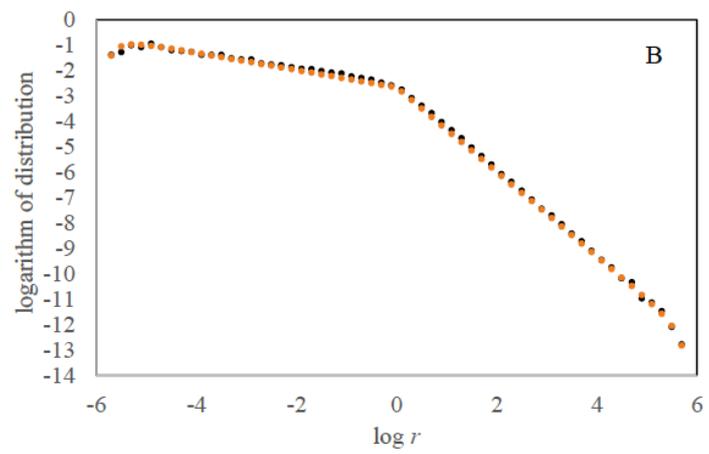

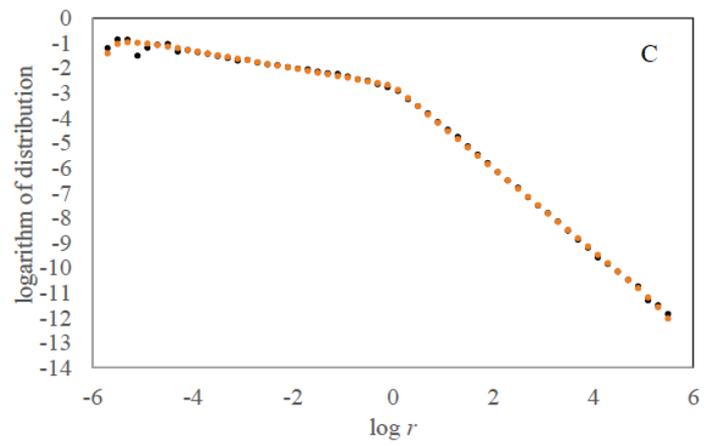

FIG. 3

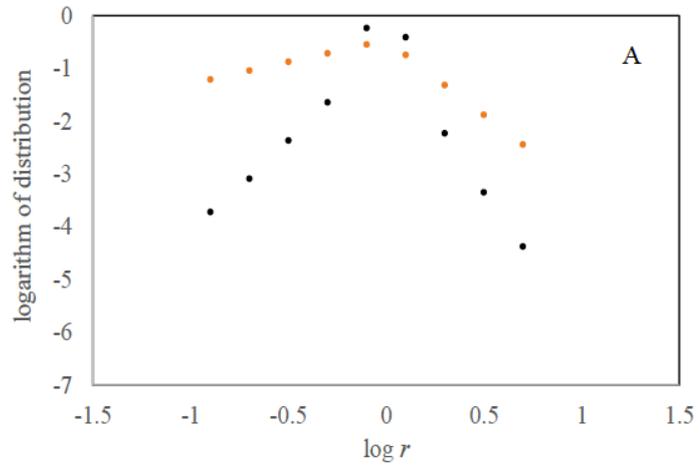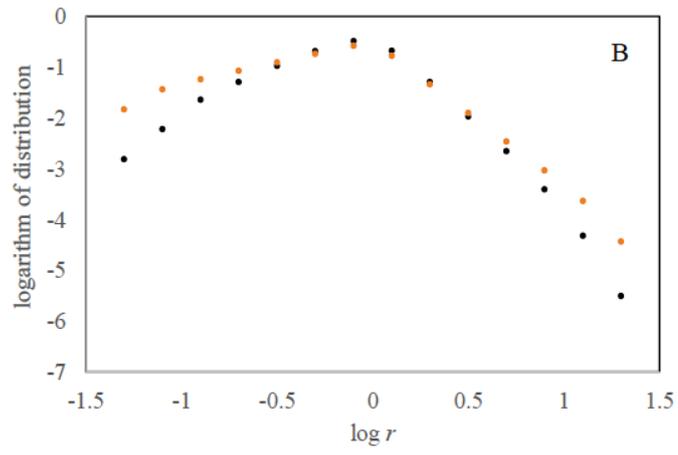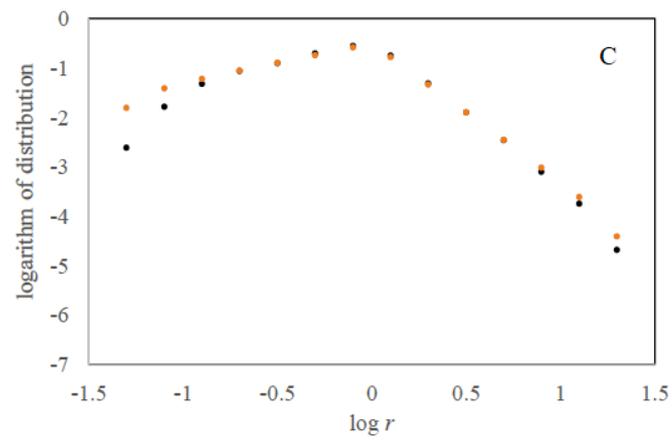

FIG. 4

# Tables

TABLE I

|             | $\varepsilon$          | $\Lambda$              | $\mathcal{N}$ |
|-------------|------------------------|------------------------|---------------|
| Seismicity-1 | $1.585 \times 10^{14}$ | $9.441 \times 10^{19}$ | 214860        |
| Seismicity-2 | $1.000 \times 10^{13}$ | $3.981 \times 10^{22}$ | 336546        |
| Brain-1     | 25.397                 | 623.590                | 162247        |
| Brain-2     | 63.297                 | 1580.464               | 233925        |

TABLE II

|  | $m = 1$ | $m = 100$ | $m = 10000$ |
|---|---|---|---|
| Seismicity-1 | | | |
| $\alpha$ | 1.66±0.01 | 1.66±0.01 | 1.65±0.01 |
| $\alpha'$ | 1.66±0.01 | 1.66±0.01 | 1.65±0.01 |
| $\alpha^*$ | 1.66 | 1.66 | 1.65 |
| Seismicity-2 | | | |
| $\alpha$ | 1.45±0.01 | 1.45±0.01 | 1.46±0.01 |
| $\alpha'$ | 1.45±0.01 | 1.45±0.01 | 1.45±0.01 |
| $\alpha^*$ | 1.45 | 1.45 | 1.45 |
| Brain-1 | | | |
| $\alpha$ | 2.82±0.09 | 2.82±0.09 | 2.79±0.09 |
| $\alpha'$ | 2.82±0.09 | 2.82±0.09 | 2.79±0.09 |
| $\alpha^*$ | 2.82 | 2.82 | 2.79 |
| Brain-2 | | | |
| $\alpha$ | 2.67±0.10 | 2.67±0.10 | 2.65±0.10 |
| $\alpha'$ | 2.67±0.10 | 2.67±0.10 | 2.65±0.10 |
| $\alpha^*$ | 2.67 | 2.67 | 2.65 |

TABLE III

|  | $m = 1$ | $m = 100$ | $m = 10000$ |
|---|---|---|---|
| Seismicity-1 | $7.91 \times 10^{-2}$ | $9.76 \times 10^{-3}$ | $9.84 \times 10^{-3}$ |
| Seismicity-2 | $6.36 \times 10^{-2}$ | $1.26 \times 10^{-1}$ | $7.33 \times 10^{-2}$ |
| Brain-1 | 2.28 | $2.59 \times 10^{-1}$ | $6.41 \times 10^{-2}$ |
| Brain-2 | 3.16 | $2.26 \times 10^{-1}$ | $4.64 \times 10^{-2}$ |

# Figure and Table Captions

FIG. 1. (A) The semi-log plot of a part of the time series of the released energy $E = 10^{11.8+1.5M}$ [J] in the unit of 1 J with respect to the numbering label (i.e., event time) $n$ of the events in the chronological order. Here, only events between $n = 405000$ and $n = 445000$ in Seismicity-1 are shown. The largest event at $n = 424588$ is the Baja California Earthquake. (B) The log-log plot of the dimensionless normalized distribution $P_{EQ}(E)$ of the released energy in the total time interval analyzed. The histogram is made by the use of bin size that gives five points in each single order of magnitude. The red straight line shows the pure power-law distribution $P_{EQ}(E) \sim E^{-1.66}$ in the interval $\varepsilon \leq E \leq \Lambda$ with $\varepsilon = 1.585 \times 10^{14}$ J and $\Lambda = 9.441 \times 10^{19}$ J [see Eqs. (1) and (2)]. The dimensionless $b$-value is $b = 0.99$, here. The number of events contained in this scaling region is $\mathcal{N} = 214860$.

FIG. 2. (A) The semi-log plot of the time series of the absolute value of the voltage difference $|V|$ [$\mu$V] in Brain-1. A single epileptic seizure indicated by the large signals between $n = 263786$ and $n = 286826$ is included. (B) The log-log plot of the dimensionless normalized distribution $P_{EEG}(|V|)$ of the voltage difference. The histogram is made by the use of bin size that gives five points in each single order of magnitude. The red straight line shows the pure power-law distribution $P_{EEG}(|V|) \sim |V|^{-2.82}$ in the interval $\varepsilon \leq |V| \leq \Lambda$ with $\varepsilon = 25.397\,\mu$V and $\Lambda = 623.590\,\mu$V [see Eqs. (1) and (2)]. The number of signals contained in this scaling region is $\mathcal{N} = 162247$.

FIG. 3. The log-log plots of the dimensionless probability distributions of the dimensionless variable $r = X'/X$ for Seismicity-1. Here, the random variables $X$ and $X'$ are the released energies of earthquakes. The red dots represent the theoretical distribution to be satisfied by a process without memory, whereas the black dots describe the real data. The histograms are made by the use of bin size that gives five points in each single order of magnitude. The interval for scaling, the values of the exponents, and the normalized squared logarithmic distance for the shifts $m = 1, 100, 10000$ are given in Tables I-III.

FIG. 4. The log-log plots of the dimensionless probability distributions of the dimensionless variable $r = X'/X$ for Brain-1. The random variables $X$ and $X'$ are the absolute values of the voltage difference. The red dots represent the theoretical distribution to be satisfied by a process without memory, whereas the black dots describe the real data. The histograms are made by the use of bin size that gives five points in each single order of magnitude. The interval for scaling, the values of the exponents, and the normalized squared logarithmic distance for the shifts $m = 1, 100, 10000$ are given in Tables I-III.

TABLE I. The values of the lower and upper thresholds, $\varepsilon$ and $\Lambda$, of the scaling regions, and the number of events and signals, $\mathcal{N}$, contained in the regions. The unit of the bounds is J for Seismicity-1 and Seicmicity-2, whereas it is $\mu\text{V}$ for Brain-1 and Brain-2.

TABLE II. The values of the dimensionless exponents in Eqs. (1), (2), and (4) for the shifts $m = 1, 100, 10000$. $\alpha$ and $\alpha'$ are evaluated from the data by the least squares method.

TABLE III. The values of the dimensionless normalized squared logarithmic distance, $d^2$, for the shifts $m = 1, 100, 10000$. The values of the exponent, $\alpha^*$, employed are given in Table II. The result on Seismicity-2 shows that the distance oscillates with respect to $m$, although global decreasing trend is confirmed.